\documentclass[10pt,twoside,leqno]{article}

\usepackage{amsmath} 
\usepackage{amssymb} 
\usepackage{subfigure} 
\usepackage{graphicx}
\usepackage{framed}
\usepackage{natbib}
\graphicspath{{figures/}}
\usepackage{float}
\usepackage[small]{caption}
\usepackage{rotating}
\usepackage{booktabs}
\usepackage{url}
\usepackage{enumerate}
\usepackage{bm}
\usepackage{nopageno}

\floatstyle{ruled}
\newfloat{algorithm}{htbp}{lop}
\floatname{algorithm}{Box}

\begin{document}

\bigskip

\begin{center}

{\Large \textbf{Robust model-based clustering with gene ranking}}

\bigskip

BY ALBERTO COZZINI$^{1}$, AJAY JASRA$^{2}$ \& GIOVANNI MONTANA$^{1}$

{\footnotesize $^{1}$Department of Mathematics,
Imperial College London, London, SW7 2AZ, UK.}\\
{\footnotesize E-Mail:\,}\texttt{\emph{\footnotesize a.m.cozzini@ic.ac.uk, g.montana@ic.ac.uk}}\\
{\footnotesize $^{2}$Department of Statistics \& Applied Probability,
National University of Singapore, Singapore, 117546, Sg.}\\
{\footnotesize E-Mail:\,}\texttt{\emph{\footnotesize staja@nus.edu.sg}}
\end{center}

\begin{abstract}
Cluster analysis of biological samples using gene expression measurements is a common task which aids the discovery of heterogeneous biological sub-populations having distinct mRNA profiles. Several model-based clustering algorithms have been proposed in which the distribution of gene expression values within each sub-group is assumed to be Gaussian. In the presence of noise and extreme observations, a mixture of Gaussian densities may over-fit and overestimate the true number of clusters. Moreover, commonly used model-based clustering algorithms do not generally provide a mechanism to quantify the relative contribution of each gene to the final partitioning of the data. We propose a penalised mixture of Student's t distributions for model-based clustering and gene ranking. Together with a bootstrap procedure, the proposed approach provides a means for ranking genes according to their contributions to the clustering process. Experimental results show that the algorithm performs well comparably to traditional Gaussian mixtures in the presence of outliers and longer tailed distributions. The algorithm also identifies the true informative genes with high sensitivity, and achieves improved model selection. An illustrative application to breast cancer data is also presented which confirms established tumor subclasses.
\end{abstract}


\section{Introduction}

Microarray gene expressions studies are routinely carried out to measure the transcription levels of an organism's genes. A common aim in the analysis of expressions measurements observed in a population is the identification of naturally occurring sub-populations. In cancer studies, for instance, the identification of sub-groups of tumours having distinct mRNA profiles can help discover molecular fingerprints that will define subtypes of disease \citep{Smolkin2003}.

Many different approaches have been suggested for partitioning biological samples, including hierarchical clustering, K-means and probabilistic models based on finite mixtures of distributions. One of the widely recognised advantages of model-based clustering lies in the fact that it explicitly accounts for the experimental noise that is typical of microarray studies \citep{Liu2010}. The gene expression measurement within each cluster are modelled as random variables drawn from a group-specific probability distribution, which is often taken to be Gaussian; several well-known parameter estimation algorithms exist and have been applied to gene expression data \citep{Qu2004, He2006, Liu2010}. 

Despite the popularity of mixture models based on Gaussian components, this particular choice of probability distributions may not always be ideal, especially in the presence of high measurement noise. A mixture model that uses Gaussian components may suffer from a lack of robustness against outliers, which in turn may lead to an inflated number of detected clusters. This is because of the the fact that additional components are needed to capture the heavy tail distribution that characterise certain groups \citep{Qu2004,Melnykov2010, Liu2010}. As an alternative to Gaussian components, Student $t-$distributions have been successfully used for robust model-based clustering in several application domains \citep{Peel2000, Liu1995}, including the analysis of gene expression data \citep{Jiao2008, Jiao2010}. The density of a Student's $t$ can achieve slower exponential tail decay, thus yielding heavier tails, and making it more robust to outliers or sampling errors \citep{Kotz2004, Peel2000}. 

The majority of algorithms for unsupervised data partitioning use all the variables to describe the objects to be clustered. However, in microarray studies, it is expected that not all the gene expression measurements will necessarily contribute equally to the identification of distinct sub-groups of samples. Even when real clusters exist and are well separated, is it often the case that only a subset of genes will have expression levels that significantly vary across groups. Failing to identify the truly informative gene expressions may yield inaccurate clustering results because the non-informative genes will mask the underlying structure of the data. The problem of {\it sub-space clustering} consists in detecting clusters that only exist in a reduced number of dimensions, and can be addressed as a variable selection problem. However, compared to supervised learning settings, such as regression and classification, variable selection in clustering is notoriously more difficult. 

Motivated by the challenges posed by the high-dimensional settings, recently there has been a burst of activity on variable selection methods applied to model-based clustering. Within the Bayesian estimation framework, most studies have adopted a specific prior that induces a sparse solution \citep{Liu2003, Tadesse2005, Yau2011}. Maximum likelihood estimation approaches achieve variable selection imposing penalties constraints on the likelihood which has the effect of shrinking some parameters to common values \citep{ Pan2007, Xie2008, Wang2008}. Since these methods rely upon the normality assumption, their performance is expected to degrade as the level of noise increases, and generally they do not provide means for ranking the genes in order of clustering importance.

In this work we propose a robust probabilistic clustering algorithm that simultaneously: (a) identifies the informative genes and ranks them by importance, and (b) discovers data clusters that emerge only when considering the expression levels of the selected genes. Our approach consists of a penalised finite mixture of Student's $t$ distribution for model-based clustering and gene selection. As noted above, the use of group-specific distributions having longer tails results in clustering assignment that are less prone to be affected by extreme or unusual observations, and facilitates the discovery of the true underlying number of clusters. Gene selection is achieved by imposing a penalty term on cluster-specific model parameters. Moreover, we propose a data resampling procedure to quantify the contribution of each gene to the clustering process and improve the inferential process of discovering the true number of clusters. 

The article is organised as follows. In Section 2 we introduce the proposed penalised finite mixture of Student's $t$ distributions, derive an EM algorithm for parameter estimation, and present a bootstrap procedure that, when combined with the BIC criterion, enhances the model selection process. In Section 3 we present results based on extensive Monte Carlo simulations which we use to assess the performance of the proposed methodology under different data generating mechanisms and in comparison to two competing algorithms. In Section 4 we discuss an application of our clustering framework to a breast cancer data set. We conclude with a discussion on potential directions of future research in Section 5.

\section{Penalised Finite Mixtures of $t$-Distributions} \label{model}

\subsection{Model}

We observe $p$ gene expression measurements on $n$ independent samples. These values are collected in $n$ vectors $\bm{y}_1, \ldots, \bm{y}_n$, which are then arranged in an $n \times p$ data matrix $\bm{y}$ and have been standardised. Our first objective is to partition the $n$ samples into naturally occurring groups.
We assume that the density of $\bm{y}$ is a mixture of $g$ components, one component for each cluster, 
\begin{equation}
        \label{genmix} g (\bm{y}|\bm{\Psi})=\prod_{j=1}^{n}\sum_{i=1}^{g}\pi_{i}\,{\it f}_{i}(y_{j}|\bm{\Theta}_{i}) 
\end{equation}
where $\bm{\Psi}= \{ \bm{\pi}, \bm{\Theta}_1, \ldots, \bm{\Theta}_g\}$ is the parameter set containing the mixing coefficients $\bm{\pi}=(\pi_1, \ldots, \pi_g)$, summing up to one, and the parameter sets $\bm{\Theta}_1, \ldots, \bm{\Theta}_g$ that fully specify the mixture components. 

Furthermore, we assume that each component of the mixture follows a multivariate Student $t-$distribution having density 
\begin{equation}
        \label{tcanonical} {\it f}({\bm y} |{\bm \mu},{\bm \Sigma},{\bm \nu}) = \frac{\Gamma (({\bm \nu} + p)/2)|{\bm \Sigma}|^{-1/2} }{\Gamma({\bm \nu}/2)(\pi{\bm \nu})^{p/2} \{1+\delta ({\bm y};\,{\bm \mu},\,{\bm \Sigma})/{\bm \nu}\}^{({\bm \nu} + p)/2}}, \nonumber
\end{equation}
where $\Gamma(\cdot)$ is the gamma function and $\delta(y;{\bm \mu},{\bm \Sigma})=(y-{\bm \mu})^{T}{\bm \Sigma}^{-1}(y-{\bm \mu})$ is the Mahalanobis squared distance. In this case, each $\bm{ \Theta}_{i}$ includes a $p$-dimensional location vector $\bm{ \mu}_{i}$, a $p \times p$ covariance matrix $\bm{ \Sigma}_{i}$ and the vector of degrees of freedom $\bm{\nu}_{i}$. The log-likelihood function can then be written as 
\begin{equation}
        \label{loglmixt} \log{L}(\bm{\Psi})= \sum_{j=1}^{n} \left[ \log \left( \sum_{i=1}^{g} \pi_{i} \, {\it f}_{i}(\bm{ y}_{j}|\bm{ \mu}_{i},\bm{ \Sigma}_{i}, \nu_{i}) \right) \right] 
\end{equation}
and the inferential problem consists in estimating $\bm{\Psi}$ from the data. In the following derivations we assume a diagonal covariance matrix, $\bm{ \Sigma}_{i}=diag(\sigma^2_{i,1},\ldots,\sigma^2_{i,p})$ with $\bm{\sigma}_i=\{\sigma_{i,1},\ldots,\sigma_{i,p}\}$ the $p$-dimensional vector of standard deviations. This assumption makes the algorithm scale well to very high dimensional setting without hindering its ability to select the truly informative variables; see for instance \cite{Xie2008}. Our second objective is to identify the gene expressions that are relevant or informative for the identification of the $g$ clusters. 

We embrace a penalised log-likelihood framework, in which the log-likelihood function \eqref{loglmixt} is replaced by 
\begin{equation}
        \label{Penloglik} \log{L}_p(\bm{\Psi}) = \log{L}(\bm{\Psi}) - h_{\bm{\lambda}}(\bm{\Theta}_1, \ldots, \bm{\Theta}_g ) 
\end{equation}
where $h(\cdot)$ is a penalty function that depends on the parameters of the mixture components as well as a regularisation parameter vector $\bm{\lambda}$. The effect of the penalty is that of shrinking the estimated parameters of the uninformative variables towards a single value that is common across all mixture components. 

For finite mixtures of Gaussian distributions, the maximisation of \eqref{Penloglik} under a penalty on the $L_{1}$-norm of the location and scale parameters has been proved to identify uninformative gene expressions \citep{Pan2007, Xie2008}. In this study we adopt a similar penalty term for our model \eqref{loglmixt}, that is 
\begin{equation}
        \label{penfun} h_{\bm{\lambda}} =\lambda_{\mu} \sum_{i=1}^{g} \sum_{d=1}^{p} |\mu_{i,d}| + \lambda_{\sigma} \sum_{i=1}^{g} \sum_{d=1}^{p} |\log \sigma^{2}_{i,d}| 
\end{equation}
where $\bm{\lambda}=(\lambda_{\mu},\lambda_{\sigma})$. According to this penalty, the log-likelihood \eqref{loglmixt} has a  penalisation that is proportional to the absolute values of the means and log-variances, where $\lambda_{\mu} \in \mathbb{R}^{+}$ and $\lambda_{\sigma} \in \mathbb{R}^{+}$ are the regularisation parameters for means and variances, respectively. 

The model is first re-written in a missing data framework, which facilitates parameter estimation \citep{Mclachlan2000}. A $g$-dimensional component-label vector $\bm{z}_j=( z_{j1}, \ldots, z_{jg} )$ is introduced to indicate the cluster membership. With probability $\pi_i$, each value $z_{i,j}$ is equal to one if the observation $y_j$ belongs to component $i$ and zero otherwise. Then $\bm{z}_j$ follows a multinomial distribution with parameters $(\pi_1, \ldots, \pi_g)$. We adopt a hierarchical representation of the Student's $t$ distribution. Conditional on $z_{i,j}=1$, $y_{j}$ follows a $t$ distribution. Introducing a second latent variable $u_j$, and using a hierarchical representation, we write 
\begin{equation}
        \label{h1} u_{i,j} \sim \text{Gamma} \left({\nu_{i} \over 2 }, {\nu_{i} \over 2 }\right) \quad y_{j} | u_{i,j} \sim \text{N}\left(\bm{ \mu}_{i},{\bm{ \Sigma}_{i} \over u_{i,j}} \right) 
\end{equation}
where $\text{Gamma}(\cdot,\cdot)$ and $\text{N}(\cdot,\cdot)$ are the gamma and the multivariate normal density functions, respectively. The complete data becomes $\bm{ y_{c}} = \{\bm{ y}_{j},\,\,\bm{ z}_{j},\,\,u_{j} \}$, and the complete data log-likelihood function can then be factorized as

\begin{equation}
        \label{penllik} \log {\mathit L}_{p}(\bm{\Psi}) = l_1(\bm{ \pi}) + l_2(\bm{ \nu}) + l_3(\bm{ \mu, \Sigma}) - h_{\lambda}(\bm{ \Theta}) 
\end{equation}
where 
\begin{eqnarray*}
        \begin{split}
                l_1(\bm{ \pi}) & = \sum_{i=1}^{g} \sum_{j=1}^{n} z_{i,j}\, \log \pi_{i} \\
                l_2(\bm{ \nu}) & = \sum_{i=1}^{g} \sum_{j=1}^{n} z_{i,j} \{ -\log \Gamma({\nu_{i} \over 2} ) + {1 \over 2}\log \Gamma({\nu_{i} \over 2} )\\
                & + {\nu_{i} \over 2} (\log u_{j} - u_{j}) - \log u_{j} \} \\
                l_3(\bm{ \mu, \Sigma}) & = \sum_{i=1}^{g} \sum_{j=1}^{n} z_{i,j} \{ -{p \over 2} \log(2 \pi) +{p \over 2} \log( u_{j}) -{1 \over 2}\log|{\bm \Sigma}_{i}| \\
                & - {1 \over 2} \, u_{j} \,({\bm y}_{j}-{\bm \mu}_{i})^{T}{\bm \Sigma}_{i}^{-1}({\bm y}_{j}-{\bm \mu}_{i}) \} 
        \end{split}
\end{eqnarray*}
and $h_{\lambda}(\bm{ \Theta})$ is as given in \eqref{penfun}.

\subsection{The EM algorithm} 

Maximum likelihood estimation (MLE) is carried out by determining the parameter values for $\bm{\Psi}$ that maximise \eqref{Penloglik}. As commonly done in similar settings, we develop an expectation-maximization (EM) algorithm \citep{Dempster1977}. At each iteration $k$ the algorithm first evaluates the expectation of the penalised log-likelihood function of the complete data conditional on $\bm{\Psi}^{(k-1)}$ and then updates the MLE of $\bm{ \Psi}$, until convergence to a local maximum.

The conditional expectation of the complete data taken with respect to the posterior probability of the latent variables is 
\begin{equation}
        \label{condexp} {\mathit Q}_{P}(\bm{ \Psi}| \bm{\Psi}^{(k-1)} ) = {\mathit E}_{\bm{ \Psi}^{(k-1)}}\{\log {\mathit L_{c}}(\bm{\Psi})|\, \bm{ y} \} - h_{\bm{\lambda}}(\bm{ \Theta}) 
\end{equation}
where $\text{E}_{\bm{\Psi}^{(k-1)}}\{\log {\mathit L_{c}}(\Psi)|\, \bm{y} \}$ is the conditional expectation of the log-likelihood. It can be noted that the penalisation term $h_{\bm{\lambda}}(\bm{ \Theta})$ does not depend on any latent variable. Using \eqref{penllik}, the conditional expectation of the log-likelihood can be decomposed as 
\begin{equation*}
        Q_{1}(\bm{ \pi}| \bm{ \Psi}^{(k-1)} ) + Q_{2}(\bm{ \nu}|\bm{ \Psi}^{(k-1)}) + Q_{3}(\bm{ \mu}, \bm{ \Sigma}|\bm{ \Psi}^{(k-1)}). 
\end{equation*}

At iteration $k$, in the E-step the expected values of the latent variables are derived from their conditional posterior distribution given $\bm{ \Psi}^{(k)}$. The conditional expectation of the indicator variable $\bm{ z}$, that is $\text{E}_{\bm{\Psi}^{(k-1)}} \left( z_{i,j}|\, \bm{ y}_{j} \right)$, is given by $$ \frac{\pi_{i}^{(k-1)} \, f_{i}(\bm{ y}_{j}|\bm{ \mu}_{i}^{(k-1)},\bm{ \Sigma}_{i}^{(k-1)}, \nu_{i}^{(k-1)})}{f(\bm{ y}_{j}|\bm{ \Psi}^{(k-1)})} \equiv \tau_{i,j}^{(k)} $$ whereas the conditional expected value of the scaling factor, is given by $$ \text{E}_{\bm{\Psi}^{(k-1)}} \left( u_{j}|\, \bm{ y}_{j},z_{i,j}=1 \right) = \frac{\nu_{i}^{(k-1)} + p}{\nu_{i}^{(k-1)} + \delta (\bm{ y}_{j};\,\bm{ \mu}_{i}^{(k-1)},\,\bm{ \Sigma}_{i}^{(k-1)}) } \equiv u_{i,j}^{(k)}. $$

Finally, the conditional expected value of the log precision factor is obtained as $$ \text{E}_{\bm{\Psi}^{(k-1)}} \left(\log u_{j}|\, \bm{ y}_{j},z_{i,j}=1 \right) = \log u_{i,j}^{(k)} + \left \{ \psi \left( \frac{\nu_{i}^{(k-1)}+p}{2} \right) - \log \left( \frac{\nu_{i}^{(k-1)}+p}{2} \right) \right \} $$ where $\psi$ is the digamma distribution, $\psi(s)= \{ 
\partial{\Gamma(s)} / 
\partial{s} \}/ \Gamma(s)$. 

In the M-step the updated estimate of $\bm{ \Psi}$ is found by maximising (\ref{condexp}) given $\bm{ \Psi}^{(k-1)}$ and given the expected values of the latent variables computed in the E-step, that is 
\begin{equation}
        \label{argmax} \bm{\bm{\Psi}}^{(k)}= \arg\max_{\bm{\psi}}{\mathit Q}_{P}(\bm{\Psi}| \bm{\Psi}^{(k-1)} ) 
\end{equation}

Since all terms in \eqref{penllik} are additive and depend on different parameters, we can solve (\ref{argmax}) by maximising each term separately. The new estimated value of $\bm{ \pi}$ is the root of the derivative of ${\mathit Q}_{1}(\bm{ \pi}|\bm{ \Psi}^{(k)})$ with respect to $\pi$. Using a Lagrange multiplier to enforce the constraint $\sum_{i=1}^{g} \pi_{i}=1$, the update for $\pi_{i}$ is $$ \pi_{i}^{(k)} = \sum_{j=1}^{n} {\tau}_{i,j}^{(k)} / n. $$

The term $Q_{2}(\bm{ \nu}|\bm{ \Psi}^{(k-1)})$ is a function of the degrees of freedom. No closed form solution is available, but the first derivative with respect to $\nu_{i}$ is smooth enough to have a straightforward numerical solution that can be found by any standard optimisation algorithm. In our implementation, we use the PORT routines in the R package {\tt nlminb}. 

The third term $Q_{3}(\bm{ \mu}, \bm{ \Sigma}|\bm{ \Psi}^{(k-1)})$ is the only one depending on parameters that are subject to regularisation. In this case we set up a constrained maximisation problem that accounts for the relevant penalty term. First, an update for the penalised mean vector $\bm{ \mu}$ is then obtained by finding the maximum of 
\begin{equation*}
        \label{mmu} \sum_{i=1}^{g} \sum_{j=1}^{n} {\tau}_{i,j}^{(k)} \,{\mathit Q}_{3j}(\bm{ \mu}_{i}|\bm{ \Psi}^{(k-1)}) - \lambda_{\mu} \, \sum_{i=1}^{g} \sum_{d=1}^{p} \,|\mu_{i,d}|. 
\end{equation*}
Despite this being differentiable with respect to $\mu_{i,d}$ only when $\mu_{i,d}\neq 0$, we can still set the derivative to zero and solve: 
\begin{equation}
        \label{opm1} \sum_{j=1}^{n} {\tau}_{i,j}^{(k)} \, { {u}_{ij}^{(k)} \over \sigma^{2}_{i,d} } ( y_{j,d}-\mu_{i,d}) - \lambda_{\mu} \,\, {\rm sign} \, ( \mu_{i,d}) = 0 
\end{equation}
while for the singular case where $\mu_{i,d}=0$ the following inequality holds true: 
\begin{equation}
        \label{opm2} \frac{\sum_{j=1}^{n} |{\tau}_{i,j}^{(k)} \, {u}_{i,j}^{(k)}y_{j,d}|}{ {\sigma}^{2}_{i,d}} \leq \lambda_{\mu}. 
\end{equation}
Combining (\ref{opm1}) and (\ref{opm2}) the updating algorithm for $\bm{ \mu_{i}}$ becomes 
\begin{equation*}
        \bm{\mu}_{i}^{(k)} = \text{sign} ( \tilde{\bm{\mu}}_{i}^{k} ) \left( | \tilde{\bm{\mu}}_{i}^{k} | - \frac{\lambda_{\mu}} {\sum_{j=1}^{n} {\tau}_{i,j}^{(k)} \, u_{i,j}^{(k)} } \bm{\Sigma}_{i}^{(k)} \right)_{+} 
\end{equation*}
where it can be seen that the unpenalised MLE of the mean, that is $$ \tilde{\bm{\mu}}_{i}^{k}= \frac{ \sum_{j=1}^{n} {\tau}^{(k)}_{i,j} {u}^{(k)}_{i,j}\bm{ y}_{j} }{ \sum_{j=1}^{n} {\tau}^{(k)}_{i,j} {u}^{(k)}_{i,j} } $$ is shrunk towards zero by an amount that increases with $\lambda_{\mu}$ and is proportional to the variance scaled by the precision factor. When $\lambda_{\mu}$ is sufficiently large, $\mu_{i,d}$ collapses to zero thus making variable $y_d$ uninformative. 

In an analogous way, the update for $\bm{ \Sigma}$ is found by maximizing: $$ \sum_{i=1}^{g} \sum_{j=1}^{n} {\tau}_{i,j}^{(k)} \,{\mathit Q}_{3j}(\bm{ \Sigma}_{i}|\bm{ \Psi}^{(k-1)}) - \lambda_{\sigma} \, \sum_{i=1}^{g} \sum_{d=1}^{p} \, |\log \sigma^{2}_{i,d}| $$ for $\bm{\Sigma}_{i}$ which is differentiable everywhere except for $\sigma_{i,d}=1$. When $\sigma_{i,d} \neq 1$ its derivative with respect to $\sigma_{i,d}$ is 
\begin{equation}
        \label{ops1} \sum_{j=1}^{n} {\tau}^{(k)}_{i,j} \,\left(-\frac{1}{2 \, { \sigma}^{2}_{i,d}} \, + \, \frac{ {u}^{(k)}_{ij} (y_{j,d} - \mu_{i,d})^{2}}{2 \, { \sigma}^{4}_{i,d}} \,\right) - \frac{ \lambda_{\sigma} \,\, {\rm sign}\, ( \log \sigma^{2}_{i,d})}{ \sigma^{2}_{i,d}} 
\end{equation}
while for $\sigma_{i,d}=1$ we have 
\begin{equation}
        \label{ops2} \left | \sum_{j=1}^{n} {\tau}^{(k)}_{i,j} \,\left(-\frac{1}{2} \, + \, \frac{ {u}^{(k)}_{ij} (y_{j,d} - \mu_{i,d})^{2}}{2} \,\right) \right |\, \leq \, \lambda_{\sigma}. 
\end{equation}
The final update is obtained combining \eqref{ops1}-\eqref{ops2}, which gives the following updates for $\bm{\sigma}^{2(k)}_i$,
$$ \left [ \frac{\tilde{\bm{\sigma}}^{2(k)}_{i}} {1+ \lambda_{\sigma} \, {\rm sign}( \bm{ c_{i}}-b_{i})/b_{i} } - 1 \right ] \, {\rm sign}(|b_{i}- \bm{ c_{i}}| - \lambda_{\sigma})_{+} +1 $$ where $\tilde{\bm{\sigma}}^{2(k)}_{i}=\bm{ c_{i}}^{k}/b_{i}^{k}$ is the unpenalized maximum likelihood estimate of the standard deviation with $$ b_{i}^{(k)}=\sum_{j=1}^{n} \tau^{(k)}_{i,j}/2, \quad \bm{c}_{i}^{(k)}=\sum_{j=1}^{n} \tau^{(k)}_{i,j} \, u^{(k)}_{i,j} (\bm{ y}_{j}-\bm{ \mu}^{(k)}_{i})^{2}/2 $$ It can be noted that, when $\lambda_{\sigma}$ is sufficiently large, $\sigma_{i,d}$ is set to be one, thus making the $d^{\text{th}}$ variable uninformative.

\subsection{Bootstrap Strategies for Model Selection} \label{boot}

Parameter estimation using the EM algorithm is carried out for a fixed number of components, $g$, and a fixed penalty vector $\bm{\lambda}$ controlling how many variables are retained as informative. The selected variables, for a given value of $\bm{\lambda}$, are collected in the set $S_{\bm{\lambda}} \subseteq \left\{ 1, \ldots, p \right\}$ having cardinality $m$. 

Both the optimal number of mixture components and level of penalisation can be found by exploring a finite number of solutions guided by the Bayesian Information Criterion (BIC). In our penalised likelihood framework, we use a modified version, $\text{BIC} = -2 \, \log\, L_{P}(\bm \Psi) + r \, \log(n)$, 
where $n$ is the number of samples and $r = g-1 + 2\, g \,m + g - q$ is the effective number of parameters once the $q=p-m$ non-informative variables are excluded from the model \citep{Pan2007}. However, the BIC criterion does not always lead to the correct choice of the best model \citep{Baek2011}. In our experience, when $m$ is very small compared to $p$ and when the densities are fat tailed we find that this criterion still prefers too complex models as some degree of over-fitting still takes place. We therefore propose a bootstrap approach that is similar to the stability selection procedure originally developed for variable selection in penalised regression \citep{Meinshausen2010}. This procedure enhances model selection, but also provides a way to rank the selected variables.  

Initially we assume that $g$ is fixed. For a given $g$, we are interested in selecting an optimal set of informative genes, which should be ranked in decreasing order of importance. We search for a value of $\bm{\lambda}$ that minimises the modified BIC criterion, and call this optimal value $\bm{\lambda^{*}}$. This search can be carried out using a grid of candidate points. Then, $B$ sub-samples of the data are randomly drawn with replacement, $\{\bm{y}^{(b)}\}_{b=1}^B$, all having size $[n/2]$. For each random sub-sample $\bm{y}^{(b)}$, we fit the penalised mixture model using the EM algorithm with the given number of components $g$ and regularisation $\bm{\lambda^{*}}$. The set of variables selected in each sub-sample is called $S_{\bm{\lambda}^*}^{(b)}$. 

An indicator variable $I_d(S_{\bm{\lambda}^*}^{(b)})$ is introduced which equals $1$ if the variable $d$ has been flagged as informative for $\bm{y}^{(b)}$, and zero otherwise. The selection probability for gene $d$ is then estimated as
$$
\hat \pi_d = \frac{ \sum_{b=1}^B I_d(S_{\bm{\lambda}^* }^{(b)}) } {B}, \quad d=1,2, \dots,p.
$$
It should be noted that, whereas a single model fit obtained with the EM algorithm using the whole data set would only provide a binary indicator labelling each variable as informative or not, the selection probabilities provide a useful metric to assess the relative importance of each gene. This is with regard to both clustering as well as for ranking purposes. All the variables having a sufficiently high selection probability are then deemed informative. A threshold on $\hat \pi_d$ could be selected to control the number of false positives, as in \cite{Meinshausen2010}, although little theoretical developments are available yet. 

Apart from enabling to rank genes, the resampling approach provides the means to improve upon the model selection process. In order to estimate the correct number of mixture components, $g$, it is common routine to compare a series of models, each one having an increasing number of components, say from $2$ to $k_{\max}$, and select the model with the smallest BIC. However, when the ratio of non-informative over informative genes is high, we have found that in practice the modified BIC still tends to overestimate the number of clusters. We address this issue by proposing the following two-step procedure. For each one of the $k_{max}-1$ models being compared, we carry out the bootstrap procedure as described above, and collect in a set of cardinality $\tilde{m}$ the informative variables selected by all models over all bootstrap replicates. In a second step, we re-fit all competing models, but instead of using all the $p$ genes, we use only the $\tilde{m}$ informative genes, where $\tilde{m}$ is usually much smaller than $p$. The selected model is the one that minimises the BIC, as usual. By initially removing the non-informative variables, and therefore the amount of noise, this simple approach reduces the bias towards selecting more complex models, and improves upon the selection of the number of components, as shown in Section \ref{Simulation}.

\section{Simulation Experiments}\label{Simulation}


The gene ranking and model selection procedures have been assessed using extensive Monte Carlo simulations. We assume a sample size of $n=200$, and $m=20$ variables informative for clustering. The number of uninformative variables, $q$, is always taken to be much higher than $m$. The informative variables are sampled from a mixture of $g$ multivariate Student's t distributions. All the uninformative variables share the same parameters and also follow a Student's t distribution.

In order to explore the effects of having fatter tails on both clustering and variable selection performance, we consider two scenarios: a low degrees of freedom case (Low DoF), in which the distributions have tails that are more pronounced compared to multivariate Gaussians, and a high degrees of freedom case (High DoF), that is approximately Gaussian. We simulate data with both two and tree components. When $g=2$, the parameters of the densities are chosen to ensure that there is roughly a $30\%$ overlap between them; when $g=3$, which we indicate as A, B and C, the parameters are chosen so that there is about $25\%$ overlap between A and B, $30\%$ overlap between A and C, and around $5\%$ overlap between B and C. In a separate experiment we specifically assess the effects of increasing the number of uninformative variables, and consider two cases: $q=200$ (low noise) and $q=2000$ (high noise), while still keeping the number of informative variables fixed at $m=20$. 

We employ three performance indicators. The clustering performance, that is the ability to identify the correct clusters, is assessed using the Adjusted Rand Index (ARI) \citep{Hubert1985}, which measures the proximity between any two partitions of the samples, and therefore quantifies how close the estimated cluster memberships are from the ground truth. This index ranges between $0$ (random assignment) and $1$ (perfect matching). The ability of the model to identify truly informative variables is summarised by the sensitivity index which is computed as the ratio of informative variables that have been correctly selected by the algorithm (true positives) over $\hat m$, the informative variables selected by the model. In order to quantify the ability of the model to exclude truly uninformative variables, we also compute a specificity index as the ratio between the number of truly negative variables over $\hat q$, the uninformative variables selected by the model. 

We compare the performance of the suggested penalised Student $t-$mixture model (PTM) with two competing clustering methods that simultaneously partition the samples and identify the relevant genes: the penalised Gaussian mixture (PGM) \citep{Xie2008} which has been implemented in R, and the sparse $K$-means algorithm (PKM) \citep{Witten2010}, which uses a lasso-type penalty to select the variables and obtain a sparse hierarchical clustering, as implemented in the R package {\tt sparcl}.

For each setting being considered, we generate $100$ independent data sets and report on Monte Carlo averages. In Table \ref{tab:DofClusPerf} we consider two and three clusters, and fit the three competing models. In all cases, the correct number of clusters is pre-specified and only the number of selected variables is learnt from the data using model selection. The modified BIC criterion is used to select the degree of penalty in both PGM and PTM models, without using the bootstrap procedure, which is evaluated separately later. For PKM, we use the built-in model selection procedures that rely on multiple permutations of the data. All methods are assessed using ARI, sensitivity and specificity indexes.

\begin{table}[th]
        \begin{center}
        \begin{tabular}{cccccccc} \toprule & & \multicolumn{3}{c}{$g=2$} & \multicolumn{3}{c}{$g=3$} \\
                DoF & Model & ARI  & Sens & Spec & ARI & Sens & Spec \\
                \midrule & PKM & 0.33  & 0.38 & 1.00 & 0.10  & 0.20 & 1.00 \\
                Low & PGM & 0.57  & 0.60 & 0.96 & 0.40  & 0.70 & 0.98 \\
                & PTM & 0.72  & 0.76 & 0.97 & 0.56  & 1.00 & 0.96 \\
                \midrule & PKM & 0.62  & 0.66 & 1.00 & 0.37  & 0.77 & 1.00 \\
                High & PGM & 1.00  & 1.00 & 1.00 & 0.71  & 1.00 & 0.99 \\
                & PTM & 1.00  & 1.00 & 1.00 & 0.71  & 1.00 & 0.99 \\
                \bottomrule 
        \end{tabular}
                \caption{Performance assessment of three competing sparse clustering methods. Data simulated with parameters $n=200$, $m=20$, and $q=2000$. The correct number of mixture components is assumed known, and variable selection is performed for each simulated data set.}
                 \label{tab:DofClusPerf}
                \end{center}
\end{table}

In the low degrees of freedom and $g=2$ case, PTM achieves the highest sensitivity index at the cost of a marginally lower specificity compared to the other two models. This ability to identify and retain the truly informative variables translates into the highest average ARI for PTM. As expected, PKM performs poorly in this case as no probabilistic model is assumed, and the model is more sensitive to extreme observations. In the $g=3$ case, the specificity of all three competing models is still comparable, but PTM achieves the highest sensitivity that leads to the best clustering performance. When the distributions are Gaussian, both PGM and PTM have similar performances, as expected in this case, whereas performance of PKM is lower, especially with three clusters.

\begin{table}[th]
                \begin{center}
        \begin{tabular}{ccccc}
                        \toprule \multicolumn{1}{c}{} & \multicolumn{2}{c} {Low DoF} & \multicolumn{2}{c} {High DoF} \\
                \multicolumn{1}{c}{} Bootstrap & $q=200$ & $q=2000$ & $q=200$ & $q=2000$ \\
                \midrule 
                 No  & $0.84$ & $0.64$ & $0.88$ & $0.82$\\
                 Yes & $0.92$ & $0.86$ & $0.98$ & $0.93$\\
                \bottomrule
        \end{tabular}
      \caption{ARI for the PTM model, with and without bootstrap. Data simulated with parameters $g=3, n=200, m=20$.}
     \label{tab:variables}
\end{center}
\end{table}

The potential improvements that can be gained from the resampling procedure of Section \ref{boot}, in terms of both variable selection and clustering performance, are summarised in Table \ref{tab:variables}. Here we consider four scenarios whereby we vary the distribution used to generate the data within each cluster as well as the number of uninformative variables. The data are sampled from a mixture of three multivariate Student $t-$distributions. The sample size is $n=200$, with $m=20$ informative variables while the number of uninformative variables $q$ is taken to be both $200$ and $2000$. After running the bootstrap procedure with a fixed probability selection threshold, $\tilde \pi=0.7$, the improved ability of the model to exclude the noise variables improves the clustering performance, as quantified by the ARI. The improvement is particularly remarkable when the distributions have longer tails. 

\begin{table}[th]
        \begin{center}
        \begin{tabular}{ccccc}
            \toprule \multicolumn{1}{c}{} &\multicolumn{2}{c} {Low DoF} & \multicolumn{2}{c} {High DoF} \\
                Bootstrap & $q=200$ & $q=2000$ & $q=200$ & $q=2000$ \\
                \midrule 
                No  & $0.5~(3.7)$ & $0.0~(4.7)$ & $0.72~(3.3)$ & $0.55~(3.75)$\\
                Yes & $0.85~(2.85)$ & $0.3~(2.35)$ & $1.00~(3)$ & $1.00~(3)$ \\
                \bottomrule
        \end{tabular}
        \caption{Percentage of correctly identified mixture components in the PTM model, with and without bootstrap. Data simulated with parameters $g=3, n=200, m=20$. The average number of clusters is in brackets.}
                \label{tab:models}  
        \end{center}
\end{table}

In Table \ref{tab:models} we explore the effects of the two-step bootstrap approach described in Section \ref{boot} for the selection of the number of mixture components. For this experiment, we use the simulated data sets used to produce the results of Table \ref{tab:variables}, where the true number of clusters is $g=3$. For each scenario being considered, we search among models having up to five clusters. We report on the percentage of times the correct number of clusters is selected by the two strategies, with and without bootstrap. As expected, both model selection strategies perform better when the distribution of the simulated variables is Gaussian. In both high and low degrees of freedom scenarios, there is a notable performance gain when using the resampling scheme especially so in particularly demanding conditions when the data have fatter tails and the number of noise variables is high. We note also that in all scenarios the average number $g$ selected, quoted in brackets in Table \ref{tab:models}, is higher for the first strategy. This evidence confirms that, by reducing the number of noise variables, the bootstrap approach alleviates the overfitting problem which is particularly important in high dimensional settings. 

\subsection{Application to Breast Cancer Data}

The proposed sparse mixture model was used for the analysis of a publicly available breast cancer data set consisting of $n=128$ early-stage tumors from those collected at Nottingham City Hospital NHS Trust between 1986 and 1992 \citep{Naderi2007, Blenkiron2007}. This cohort of tumours is representative of the demographics of breast cancer and the majority of patients were post-menopausal. Microarray data recording the expression levels of $p=36,939$ probes are available in the ArrayExpress database (www.ebi.ac.uk/arrayexpress) under accession number E-TABM-576. Our analysis was unsupervised and aimed at simultaneously detecting clusters and ranking genes based on their contribution to the cluster assignments.

\begin{figure}[th] 
                \begin{center}
        \includegraphics[scale=0.6]{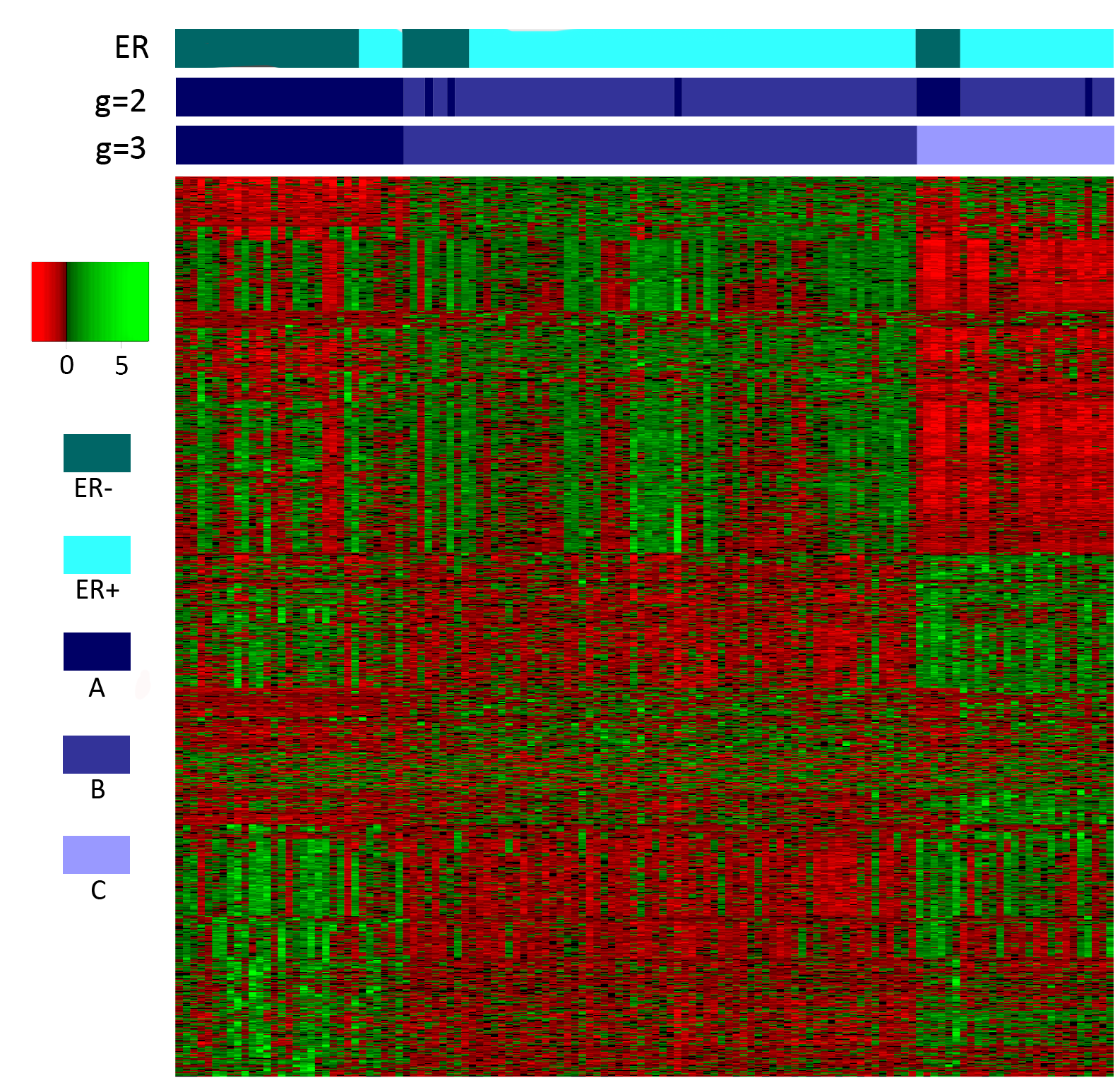}
        \vspace*{8pt}
        \caption{Heatmap showing the expression levels in $1128$ the top-ranked genes (rows) in the $128$ tumors (columns).} \label{fig:heatmap} 
\end{center}
\end{figure}

Following the model selection procedure detailed in Section \ref{model}, we inferred the number of clusters by exploring up to five possible clusters. For this data set, the minimum modified BIC supports a $3$-component mixture. The resampling approach was also used to estimate the selection probability of each probe. The heatmap of Fig. \ref{fig:heatmap} shows the $1128$ probes with selection probability at least as large as $0.7$ (about $3\%$ of the total), and their expression patters across the $128$ tumors. In order to interpret these clustering allocations, we explored the correlation with the clinical sub-groups induced by the Estrogen Receptor (ER). Approximately $80\%$ of human breast carcinomas present an estrogen receptor $\alpha$-positive (ER+) disease, with ER+ breast cancers responding well to therapies and ER-negative tumours being more resistant. ER status is an essential determinant of clinical and biological behaviour of human breast cancers and it is well known that the major molecular features of breast cancer segregate differently according to ER status \citep{Schneider2006}. With $g=2$, our analysis finds clusters that overlap with ER-positive and ER-negative tumours; these results are consistent with previous findings on independent data sets \citep[for instance]{Veer2002}. In this case, $82.5\%$ of all ER- tumors fall into cluster A and the remaining ones into cluster B. When an additional cluster is added ($g=3$), we observe a split of the ER+ dominated cluster B into two groups: $81\%$ of the samples in cluster A are still ER-, whereas the majority of samples in B and C are ER+ ($88\%$ and $76\%$, respectively). The ER status as well as the cluster assignments for both $g=2$ and $g=3$ are reported at the top row of Fig.\ref{fig:heatmap}. 

Several genes that have been selected with high selection probability by PTM had been previously included in independent ER signatures \citep{Abba2005,Thakkar2010}.
These genes include (with selection probabilities): FOXA1 (0.935), NTN4 (0.906), CSNK1A1 (0.892), STC2 (0.885), SF3B3 (0.863), MMP12 (0.860), ITGB7 (0.845), CA12 (0.831), MLPH (0.827), GATA3 (0.824), NAT1 (0.802), GABRP (0.781), DUSP4 (0.745), NUTF2 (0.777), XBP1 (0.756), SLC43A3 (0.752), PLAT (0.727), ESR1 (0.727) and AARS (0.713). 
Among the genes with highest selection probability, FOXA1 has very recently been found to be a key determinant of ER function and endocrine response \citep{Hurtado2011}. Moreover, FOXA1, GATA3 and ESR1 have also been found to be associated to ER status in an analysis of invasive ductal carcinoma \citep{Schneider2006}. Genetic variants immediately upstream of ESR1 have been linked to breast cancer risk. Very recently, \cite{Dunbier2011} found that three open reading frames within this region are tightly co-expressed with ESR1, and investigated the function of these three genes: C6ORF97, C6ORF96 and C6ORF211. Their findings suggest that the genes could contribute to the phenotype associated with ER positivity. In addition, they may be involved in the mechanism by which genetic variation in this region of the genome contributes to breast cancer susceptibility. These three genes have also been found to have high selection probabilities in our ranking (0.770, 0.713 and 0.709, respectively). 

\begin{figure}[!h] 
    \centerline{\includegraphics[scale=0.5]{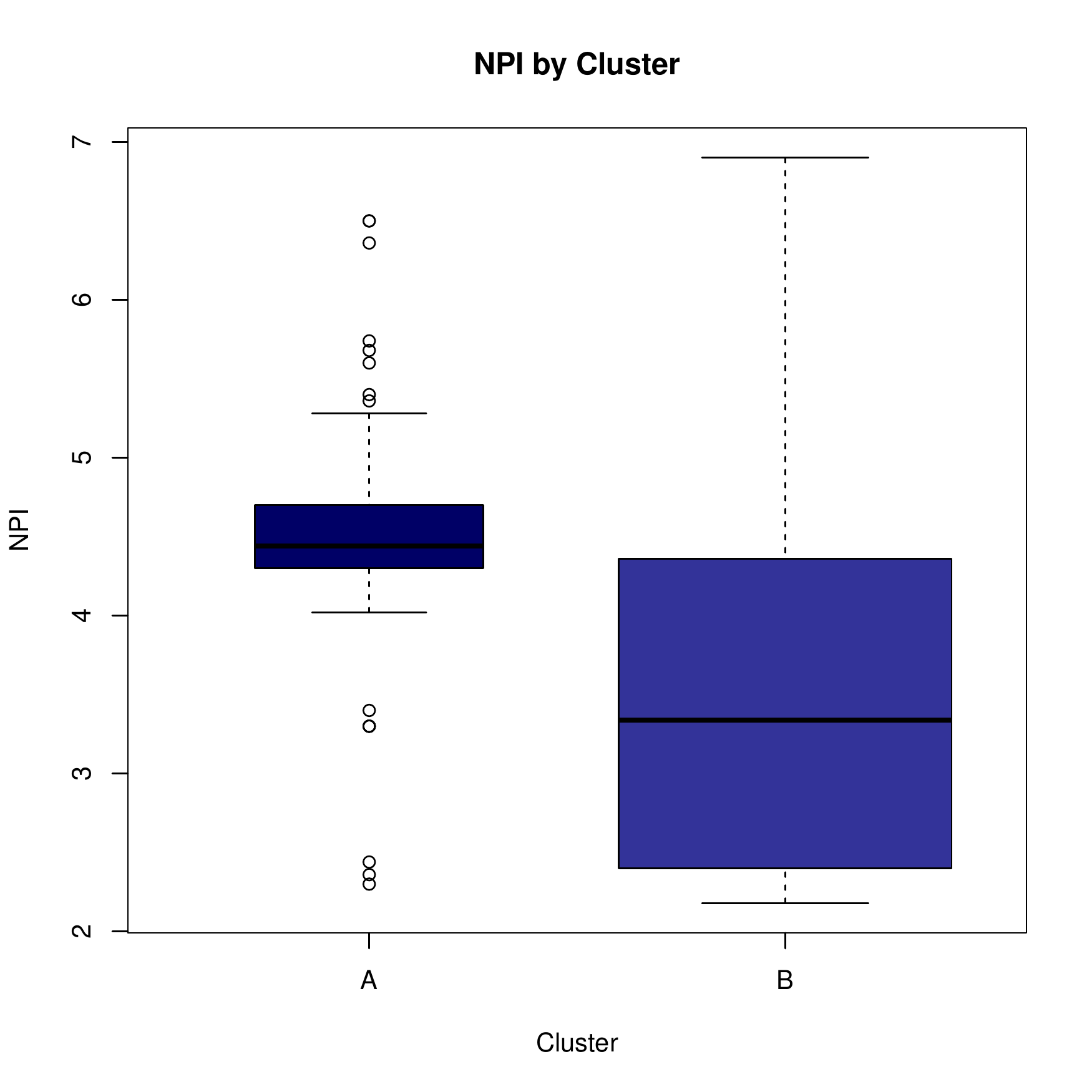}}
    \caption{NPI by cluster using two components.}
    \label{fig:NPI2}
  \end{figure}

\begin{figure}[!h] 
    \centerline{\includegraphics[scale=0.5]{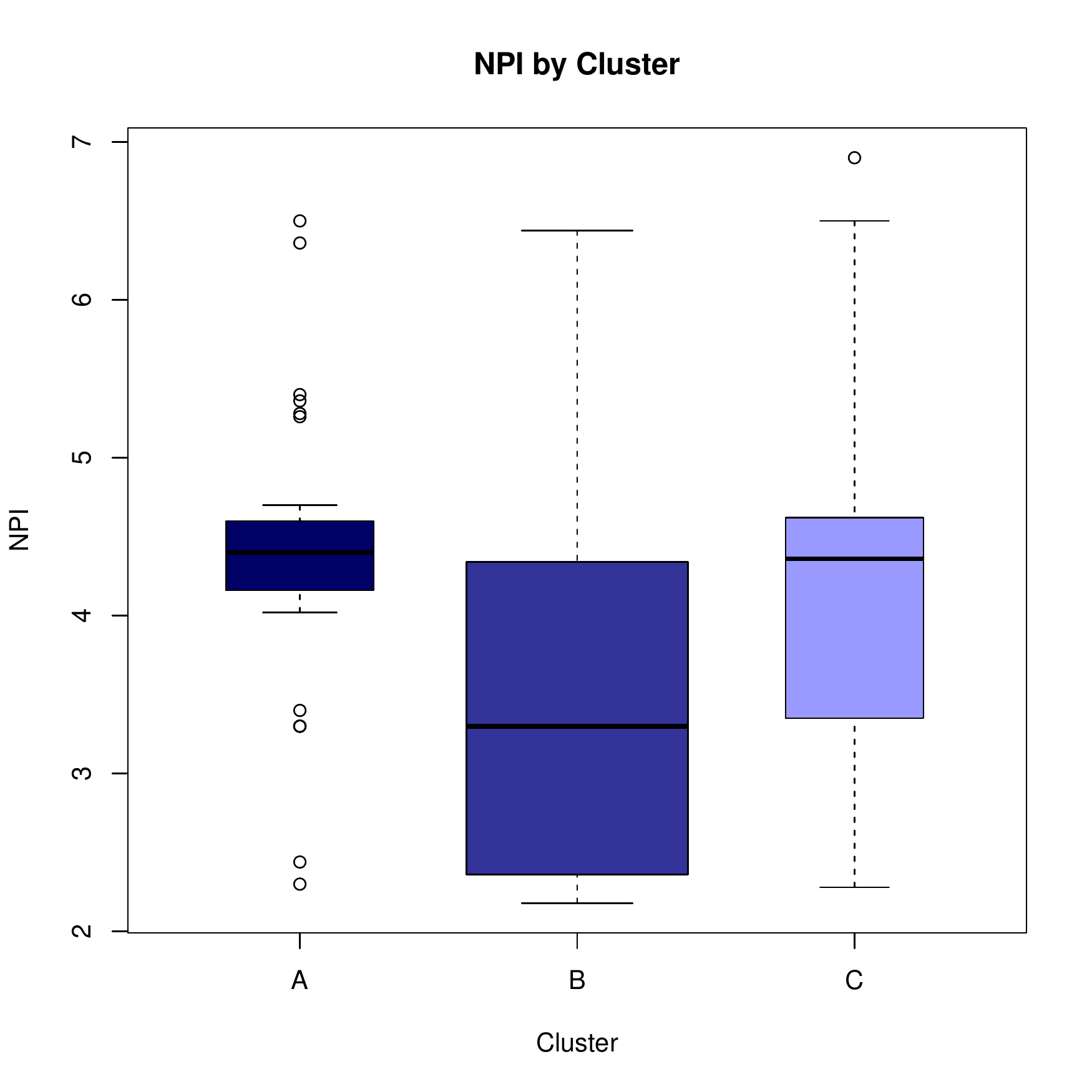}}
    \caption{NPI by cluster using three components.}
    \label{fig:NPI3}
  \end{figure}

We further investigate whether the three clusters may indicate clinically distinct subgroups of patients. We report on the frequency distribution of the Nottingham Prognostic Index (NPI) in the clusters detected by the mixture model, using both two and three mixtures components. This index uses pathological variables, such as tumor size and grade, to generate a prognostic score for each patient that is predictive of outcome \citep{Callagy2006}. When using two components (see Fig \ref{fig:NPI2}), the NPI present a different distribution in the two clusters, which are representative of ER+ and ER- tumors. When using three mixture components (see Fig \ref{fig:NPI3}), we find that patients in subgroup C have a statistically significantly higher NPI index compared to those in B, and are more similar to the ER- dominated cluster A, for which response to therapy is known to be less favourable. 

Different studies have tried to identify breast carcinomas subtypes beyond the simple classification between positive and negative ER status \citep{Sorlie2001, Calza2006, Sotiriou2009, Nicolau2011}. The subtypes varies across studies and there is no clear consensus on the gene expression signatures that identify each group. In order to better interpret our results, we have compared our proposed clustering based on the genes selected by our penalised mixture model (PTM) to the clustering obtained by using only the $87$ genes identified as informative by \cite{Calza2006} in their study of $412$ breast cancers patients from Stockholm and Uppsala, Sweden. Their gene expression signatures were originally proposed by \cite{Sorlie2003} and consisted of approximately $500$ genes whose expression varied the least in successive samples from the same patient's tumor but which showed the most variation among tumors of different patients. Genes were median-centered and subtypes were isolated using an average-linkage hierarchical clustering algorithm based on uncentered correlation distance metric.

\begin{figure}[!h] 
  \centerline{\includegraphics[scale=0.5]{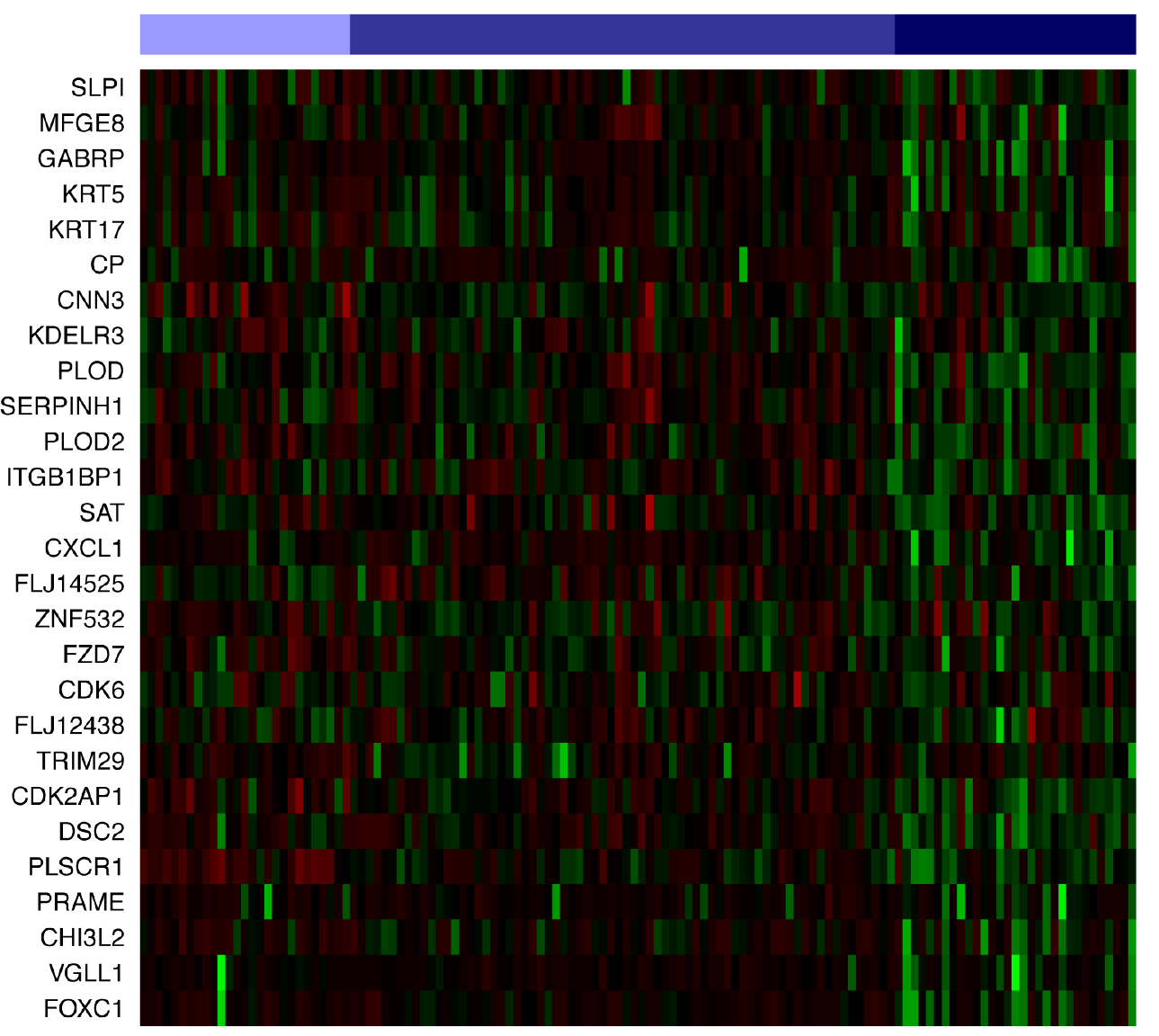}}
  \caption{The Basal signature of \cite{Calza2006} applied to the Nottingham City Hospital NHS Trust cohort identifies a Basal-like subtype that overlaps with Cluster A obtained by PTM.}
  \label{fig:heatmapba}
\end{figure}

The heatmap in Figure \ref{fig:heatmapba} shows the gene expression patterns in our cohort of $128$ samples using the Basal-like signature provided by \cite{Calza2006}. Our cluster A contains all the samples in which the marker genes are highly expressed, and is characterised by an higher percentage of ER- cases, as expected for this subtype. The clinical prognosis associated to cluster A is also in line with what has been previously reported in the literature whereby patients of this group show an above average tumor size, higher frequency of grade three carcinomas and are less responsive to therapy.

\section{Discussion and Conclusions}

In this work we have proposed a penalised finite mixture of Student $t-$distributions which performs gene selection and data clustering in a penalised maximum-likelihood estimation framework, fitted via the EM algorithm. A novel bootstrap procedure for model selection and variable ranking has also been developed. Using simulated data under various different settings, we have shown that the proposed methods are expected to be particularly beneficial in the presence of many uninformative gene expressions and fat tailed distributions. 

The algorithm has been applied for the analysis of a large data set consisting of breast cancer tumors. Our results confirm the importance of top ranked genes in the their role of differentiating between ER+ and ER- samples. Moreover, there is an indication that the selected genes may be important to define a clinically important subtype of E+ enriched tumors, which warrants further investigation.

As an extension to this work, we can consider Bayesian estimation.
As is consistent with the literature on the lasso, one can easily place a Bayesian interpretation on our penalised mixture of Student $t-$distributions.
Given the natural probabilistic nature of selecting $g$ in the Bayesian framework, it is of interest to develop advanced simulation-based estimation techniques that are required for our model.

\bibliographystyle{natbib_abbr} 
\bibliography{BioArticleEdited}



\end{document}